\documentclass[aps,pra,superscriptaddress,amsmath,amssymb,twocolumn,nofootinbib,showpacs]{revtex4-1} 
\pdfoutput=1

\usepackage{graphicx}
\usepackage{dcolumn}
\usepackage{bm}
\usepackage{bbold}
\usepackage{nonfloat}
\usepackage{color}

\newcommand{\ket}[1]{\ensuremath{|#1\rangle}}

\definecolor{pink}{rgb}{1,0.16,0.64}
\definecolor{brown}{rgb}{0.65,0.16,0.16}
\definecolor{purple}{rgb}{0.5,0,0.5}
\definecolor{grey}{rgb}{0.3,0.3,0.3}
\definecolor{green2}{rgb}{0,0.5,0}

\begin{document}

\preprint{PRE/005}

\title{Assessing quantum error correction:\\fidelity and entanglement measures with application to photonic codes}

\author{Ricardo Wickert}
 \email{ricardo.wickert@mpl.mpg.de}
\affiliation{Optical Quantum Information Theory Group, Max Planck Institute for the Science of Light, G\"unther-Scharowsky-Str. 1/Bau 26, 91058 Erlangen, Germany}
\affiliation{Institute of Theoretical Physics I, Friedrich-Alexander Universit\"at Erlangen-N\"urnberg, Staudttr. 7/B2, 91058 Erlangen, Germany}

\author{Peter van Loock$^{1,2,}$}%
\affiliation{Institut f\"ur Physik, Johannes Gutenberg-Universit\"at Mainz, Staudingerweg 7, 55128 Mainz, Germany}

\date{\today}



\begin{abstract}
By interpreting the well-known, qualitative criteria for the existence of quantum error correction (QEC) codes by Knill and Laflamme from a quantitative perspective, we propose a figure of merit for assessing a QEC scheme based on the average fidelity between codewords. This enables us to quantify the performance of a protocol as a whole, including errors beyond the correctable set. Various examples are calculated for photonic qubit codes dealing with the experimentally relevant case of photon loss, demonstrating the advantages of the new measure. In the context of continuous-variable QEC, our codeword-overlap measure can be used to reproduce, in a different way with no need for calculating entanglement measures, a previous result concerning the impossibility of improving transmission of Gaussian states through Gaussian channels via Gaussian operations alone.
\end{abstract}

\maketitle

\section{Introduction}
Operating a device or scheme in the microscopic domain places formidable demands on the purity and stability of the diverse systems involved, augmented by fundamental limitations of Quantum Mechanics \cite{NoCloning}. In this context, ingenious mechanisms to avoid the undesirable effects of decoherence play a fundamental role in many areas of Quantum Information Processing (QIP), most notably, enabling large-scale, fault-tolerant quantum computation \cite{ShorFT} and the communication of quantum bits across large distances \cite{QREncoding}. Such mechanisms typically rely on two different solutions, namely, those provided by  Quantum Error Correction (QEC) \cite{gottesmanphd}  and teleportation \cite{BennettTeleport} combined with Entanglement Distillation or Entanglement Purification Protocols (EPPs) \cite{duerbriegel}.

As these techniques mature over a multitude of implementations, with many advanced protocols already finding applications in real-world conditions, the necessity to define suitable measures to compare and rank different schemes becomes of uttermost importance. In particular, the need arises to find quantities which can be easily computed for the widening range of experimentally accessible states currently used in QIP. 
Two main measures, in different variants, currently share widespread acceptance: the entanglement degree is traditionally seen as the golden measure when defining the 'quantumness' of a channel, and is the benchmarking criterion of choice when evaluating EPPs and quantum memory devices \cite{Nathan}. 
Conversely, in the realm of QEC, fidelity measures, in particular, the worst-case fidelity \cite{KLCriteria} or entanglement fidelity \cite{schumacher},  have historically been employed. The connection between these approaches has been investigated in different regimes \cite{Bennett,duerbriegel}, but it is not always a straightforward one: Naively, one could expect codes which improve such fidelities to lead to higher safeguarded entanglement, which is indeed true in certain cases; nevertheless, there are examples \cite{bjork,DebbieApprox,EberlyNew} where a code will not safeguard any entanglement, up to and including causing entanglement sudden death (ESD) \cite{eberly}, while still improving input-output or entanglement fidelities under certain conditions. 

In this contribution, we propose a new measure based on the average fidelity between codewords of a given alphabet, emphasizing that it satisfies certain desirable features. We present initially a motivating argument starting from well-known criteria for the existence of QEC codes, and proceed with examples comparing the entanglement (as measured, in the case of qubits, by the concurrence \cite{Wooters}) to our average codeword fidelity for a variety of photonic schemes, with particular attention given to codes designed to protect a logical qubit against amplitude damping \cite{Yamamoto}. In the spirit of \cite{scottprobfail}, we go beyond the analysis traditionally restricted to the correctable error set and establish a comparison method for general schemes. However, here we impose no limitations to a code's distance or a priori assumptions on the operating regime \cite{Aschauer} (correction or detection modes \cite{ashik}). Finally, in the context of continuous-variable QEC with the logical states being infinite-dimensional, we employ our new measure to re-obtain a known result \cite{CerfNoGo} concerning the impossibility of performing QEC for Gaussian signals and channels when restricted to a Gaussian toolbox.

\section{Error Correction Conditions}
Consider a completely positive and trace-preserving map $\mathcal{E}$ corresponding to a - possibly noisy - transmission/evolution channel. Its action on an arbitrary state $\varrho$ is given by
\begin{align}
    \mathcal{E}(\varrho) = \sum_{k} A_{k}^{} \, \varrho \, A_{k}^{\dagger} \; ,
    \label{KrausEq}
\end{align}
dubbed the Kraus operator-sum representation \cite{KrausBook}. Based on this channel decomposition, the Knill-Laflamme (K-L) criteria \cite{KLCriteria,BKK} establish that an error correction map $\mathcal{R}$ satisfying $\mathcal{R}\circ\mathcal{E}=\mathcal{I}d$ exists provided it holds that
\begin{align}
\langle \chi_i | A^{\dagger}_{k}
A_{l}  | \chi_j \rangle
= \delta(i-j) \lambda_{k,l} \; ,
\label{KLConds}
\end{align}
where $|\chi_i\rangle$ are the codewords from a given input alphabet, and the $\lambda_{k,l}$ define how different errors skew the codespace; in a simplified interpretation, it requires that different codewords remain orthogonal after the action of the channel, and that different errors must effect the same deformation across the input alphabet. Exact satisfiability of the above can only be achieved in certain scenarios \cite{MintertNew}; at the same time, allowing for small deviations in the orthogonality and deformability requirements - or, alternatively, for $\mathcal{R} \circ \mathcal{E}$ to be close, but not necessarily equal, to the identity \cite{BenyApprox2,barnumknill} -  enables one to obtain more efficient codes \cite{DebbieApprox,BenyApprox}. The discrepancies may be employed to bound the  entanglement fidelity obtained with the use of the code \cite{leungstudent}.

Before proceeding, a few notes are in order. First, in achieving the codewords, an encoding step is implicitly assumed, and, for notation purposes, it is condensed together with the transmission map into a single channel \footnote{Explicitly, this means that the initial signal Hilbert space is extended, typically by adding a sufficient set of ancilla states and applying a global encoding unitary on the signal-ancillae system. The total system is then subject to a global channel transmission, which is usually composed of the original channel acting individually and independently on the signal and ancilla subsystems. In the optical setting, enlarging the Hilbert space may mean either adding extra modes or allowing for higher photon numbers with the same number of modes. In the latter case, there are no additional channels that must act on auxiliary modes.}; in the case of no encoding/direct transmission, one can simply regard the encoding as the identity map, $\mathcal{I}d$. Furthermore, we note $\delta(i-j)$ corresponds to the more general case of a continuous alphabet; in the case of qubits (or other discrete-variable codewords), this should correspond to a Kroenecker delta, $\delta_{ij}$.

Motivated by the idea of employing a primarily qualitative criteria as a potentially \textit{quantitative} measure \cite{QuantitativeWitness}, we will explore taking into account violations from the exact satisfiability of Eq.~(\ref{KLConds}) (instead of neglecting such violations up to a certain order in the channel parameters \cite{DebbieApprox}). Deviations from Eq.~(\ref{KLConds}) can be broken down in two qualitatively different types: first, violations of $\delta(i-j)$, which lead different codewords to overlap, and thus reduce the distinguishability between the input alphabet; second, departures from $\lambda_{k,l}$, which in the non-violated case is strictly independent of the codewords, but in the violated case may affect different codewords in unequal manner (and hence may deform a superposition of codewords). Obviously this latter deformation can only take place when the prior effect is also present, but the converse is not necessary \footnote{More precisely, when taking each channel Kraus operator separately like in Eq.~(\ref{KLConds}), of course each Kraus effect can affect different codewords differently, even though their overlap remains unaffected. An example is the amplitude damping channel acting on a single-rail qubit basis, for which we have $\langle 0 | A^{\dagger}_{0}
A_{0}  | 0 \rangle = \lambda_{0,0} = 1$, $\langle 1 | A^{\dagger}_{0}A_{0}  | 1 \rangle = \lambda_{0,0} = 1 - \gamma$, but also $\langle 0 | A^{\dagger}_{0}A_{0}  | 1 \rangle = 0$ (see later, Eq.~(\ref{amplitudedampingkrausoperators})). In other words, the length of vector $| 1 \rangle$ is reduced for any non-zero $\gamma$ and that of vector $| 0 \rangle$ remains unity, while they would perfectly maintain their orthogonality. However, note that by taking into account a second Kraus effect, which together with the first Kraus effect forms a trace-preserving qubit channel, the deformation is indeed accompanied by a reduction of the codeword distinguishability: $\langle 0 | A^{\dagger}_{0}A_{1}  | 1 \rangle = \sqrt{\gamma}$. Our codeword overlap measure, as defined in Eq.~(\ref{fidoverlap}) for the whole trace-preserving channel, therefore captures both deformation and non-orthogonality at the same time (see below).}. We'll dub codes (i.e., encoding together with transmission channels) in which the entire alphabet is affected uniformly as ``non-deformable", whereas those with codeword-dependent skewness will be called ``deformable".

Non-deformable codes will not preclude the distinguishability to decrease, neither prevent different pairs from suffering varying deformations: the label only guarantees, for any two orthogonal states, the overlap between the original and the resulting states to be the same. 

In the case of experimentally relevant amplitude damping channels, direct transmission of the $|0\rangle$ and $|1\rangle$ states (taken here as the occupation numbers of a bosonic mode like in so-called single-rail encoding) yields a deformable code: While the vacuum state is unaffected, the single-excitation $|1\rangle$ is taken to $(1-\gamma)|1\rangle\langle1| + \gamma|0\rangle\langle0|$. That the conjugate basis would be uniformly affected does not alter the classification of the code; in fact, for every deformable code there exists at least one superposition in which the deformations are equal. However, considering the dual-rail encoding, $|0\rangle_{L}=|0\rangle|1\rangle$ and $|1\rangle_{L}=|1\rangle|0\rangle$ - and assuming, of course, equal dampening in both rails (modes) - one finds again a non-deformable encoding, as \textit{any} superposition of the codewords will result in an equal, global reduction of the length of the logical state vector.


With the above considerations in mind, one could conceive employing the amount by which the criteria in Eq.~(\ref{KLConds}) have been violated as a measure to rank different channels.  We aim at a measure which, instead of classifying codes by the number and kind of errors it is capable of handling, should deliver information precisely about the operators the scheme is not capable of correcting, quantifying just how much it causes the input alphabet to skew and overlap. To this purpose, we construct the \textit{codeword overlap} of a map by considering the output fidelities obtained from a pair of orthogonal input states, and then averaging this quantity over the entire input alphabet:
\begin{align}
\label{fidoverlap}
F^{CW} = \int dQ \operatorname{Tr} \left[\sqrt{\sqrt{\rho_Q} \rho_{\tilde{Q}} \sqrt{\rho_Q}}\right] \; .
\end{align}
Here, $\rho_Q$ and $\rho_{\tilde{Q}}$ correspond to the outputs originating from a pair of orthogonal input states $|Q\rangle$ and $|\tilde{Q}\rangle$, with $|Q\rangle \perp |\tilde{Q}\rangle$. The above loosely corresponds to an integration of Eq.~(\ref{KLConds}), with two important distinctions: first, since the performance of a code should be basis-independent, we cover the surface of the Bloch sphere, taking each pair of diametrically opposed states as possible codewords; second, instead of acting with the Kraus operators individually, we consider the full channel's effect on the codewords, which enables it to be employed even in channels were a decomposition in form of Eq.~(\ref{KrausEq}) is not known.


When considering encodings for spaces larger than qubits (\textit{e.g.}, the qudit codes proposed in \cite{Yamamoto}), one should per -- doneform the integration considering opposing states in the surface of the corresponding Bloch hypersphere \cite{BlochHypersphere}.\footnote{The form above is of course not exactly convenient when considering such encodings of higher dimensions. Already in the case of a qutrit ($d=3$) this is easily seen: an arbitrary logical state has the general form $a\ket{0}_L + b\ket{1}_L + c\ket{2}_L$, and a possible codeword is $\ket{0}_L$, which has infinitely many states orthogonal to it (for instance, any state of the form $b\ket{1}_L + c\ket{2}_L$ with $b^2+c^2=1$). Thus, in order to calculate the measure, for every $\ket{Q}$ state, one has to consider the average fidelity to \textit{each} of the possible $\ket{\tilde{Q}}$ orthogonal states.} This, however, precludes a straightforward generalization for continuous-variable (CV) encodings; nevertheless, one may consider only a truncated set, i.e., a qudit encoding where $d \rightarrow \infty$.

\section{Photon-loss qubit codes}

Here we consider the ubiquitous binary logical basis, that is, with the information encoded in orthogonal states $|0\rangle_L$ and $|1\rangle_L$. The transmission of such states via optical fibers corresponds generally to a lossy process, with the fiber absorbing or ``losing" photons along its length. This corresponds to a Gaussian channel of particular relevance, dubbed an \textit{amplitude damping} channel. It is characterized by a loss parameter, $\gamma$, and described, in the Kraus representation of Eq.~(\ref{KrausEq}), by an infinite sum ($k=0...\infty$) of operators of the form
\begin{align}
\label{amplitudedampingkrausoperators}
A_k = \sum_{n=k}^\infty \sqrt{\binom{n}{k}} \sqrt{(1-\gamma)^{n-k} \gamma^k} |n-k\rangle \langle n| \; .
\end{align}

Different encodings have been designed to protect qubits against the errors caused by these operators, and are presented below to illustrate the usage of the new measure. The first code considered is the aforementioned dual-rail encoding, which provides \textit{detection} capabilities only:
\begin{align}
|0\rangle_L = |0 1\rangle \; \mbox{,} \quad |1\rangle_L = |1 0 \rangle \; .
\end{align}
Whenever both modes are found to be in the vacuum state, the information has to be transmitted anew (error detection mode) or, after the decoding step, the output is replaced by the mixed qubit $\rho = \frac{1}{2} |0\rangle\langle0| + \frac{1}{2} |1\rangle\langle1|$ (error correction mode, which will be considered here).

The ubiquitous three-qubit repetition code,
\begin{align}
|0\rangle_L = |0 0 0\rangle  \; \mbox{,} \quad |1\rangle_L = |1 1 1 \rangle \; ,
\end{align}
is capable of correcting any single bit-flip error, and can also be employed against amplitude damping errors, assuming photon loss to be a particular case of bit-flip, with a highly asymmetrical behaviour in which only one of the logical states is affected.

Codes can also be constructed by exploring higher occupations of the bosonic modes. For instance, in \cite{Yamamoto}, the authors develop the following encoding, capable of correcting the loss of up to one quanta to the environment:
\begin{align}
|0\rangle_L = \frac{|40\rangle + |04\rangle}{\sqrt{2}}  \; \mbox{,} \quad |1\rangle_L = |22 \rangle \; .
\end{align}

Allowing for codes which do not exactly satisfy Eq.~(\ref{KLConds}) enables one to achieve more economical encodings. One such was proposed by Leung \textit{et al}. \cite{DebbieApprox},
\begin{align}
|0\rangle_L = \frac{|0000\rangle + |1111\rangle}{\sqrt{2}}  \; \mbox{,} \quad
|1\rangle_L = \frac{|0011\rangle + |1100\rangle}{\sqrt{2}} \; ,
\end{align}
which, assuming the damping constant $\gamma$ be kept low, is capable of correcting the loss of one excitation to $O(\gamma^2)$.

Finally, the usage of infinite-dimensional carriers is one that offers numerous advantages in the realm of QIP \cite{LloydBraunstein,RalphMilburn}. One may thus consider the encoding of qubits on coherent states, which are innately non-orthogonal:
\begin{align}
|0\rangle_L = |-\alpha\rangle  \; \mbox{,} \quad |1\rangle_L = |\; \alpha\rangle \; .
\end{align}
Here, the encoding can be seen as a channel which irreversibly reduces the distinguishability; in this case, a state is no longer orthogonal to the state found on the diametrically opposing point of the Bloch sphere, with the exception of those lying in the equator. Error correction codes have also been developed in this regime, such as the protocol in \cite{GVR}, which effects, up to a normalization, 
\begin{align}
|-\alpha\rangle \rightarrow \left( |-\alpha\rangle + |\alpha\rangle \right)^{\otimes N}  \; \mbox{,} \;
|\; \alpha\rangle \rightarrow \left( |-\alpha\rangle - |\alpha\rangle \right)^{\otimes N} \; ,
\label{coherentencoding}
\end{align}
capable of correcting $\lfloor \frac{N-1}{2} \rfloor$ amplitude-damping induced phase-flip errors. We note, however, that not only the transmissivity of the channel contributes to this scheme's performance: also the size of the superposition, which regulates the amount of overlap in the input alphabet, affects the transmission characteristics (see \textit{e.g.} \cite{OurPaper,OurFuturePaper}).

We proceed by calculating the average codeword overlap for each of the above encodings and comparing it to a different figure of merit, namely, the amount of entanglement preserved after employing the protocol at hand to transmit one half of a two-qubit maximally entangled state (further details are provided in Appendix A). The measures are computed as a function of the channel loss parameter $\gamma$, in the case of the discrete-variable encodings, or for a fixed channel transmissivity, as a function of the coherent-state superposition size $|\alpha|$.


\begin{figure}[t!]
\includegraphics[scale=0.5]{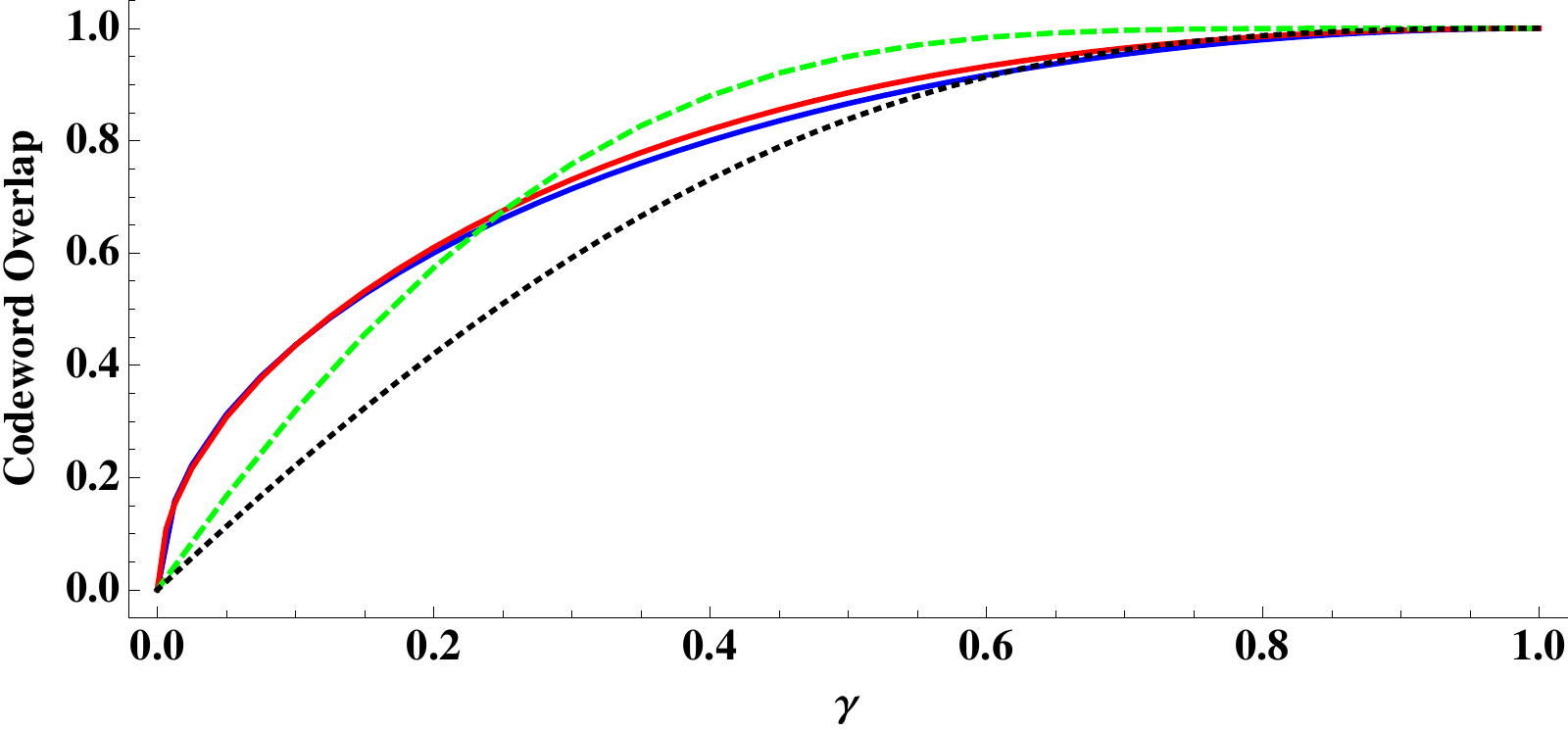} \\
\includegraphics[scale=0.5]{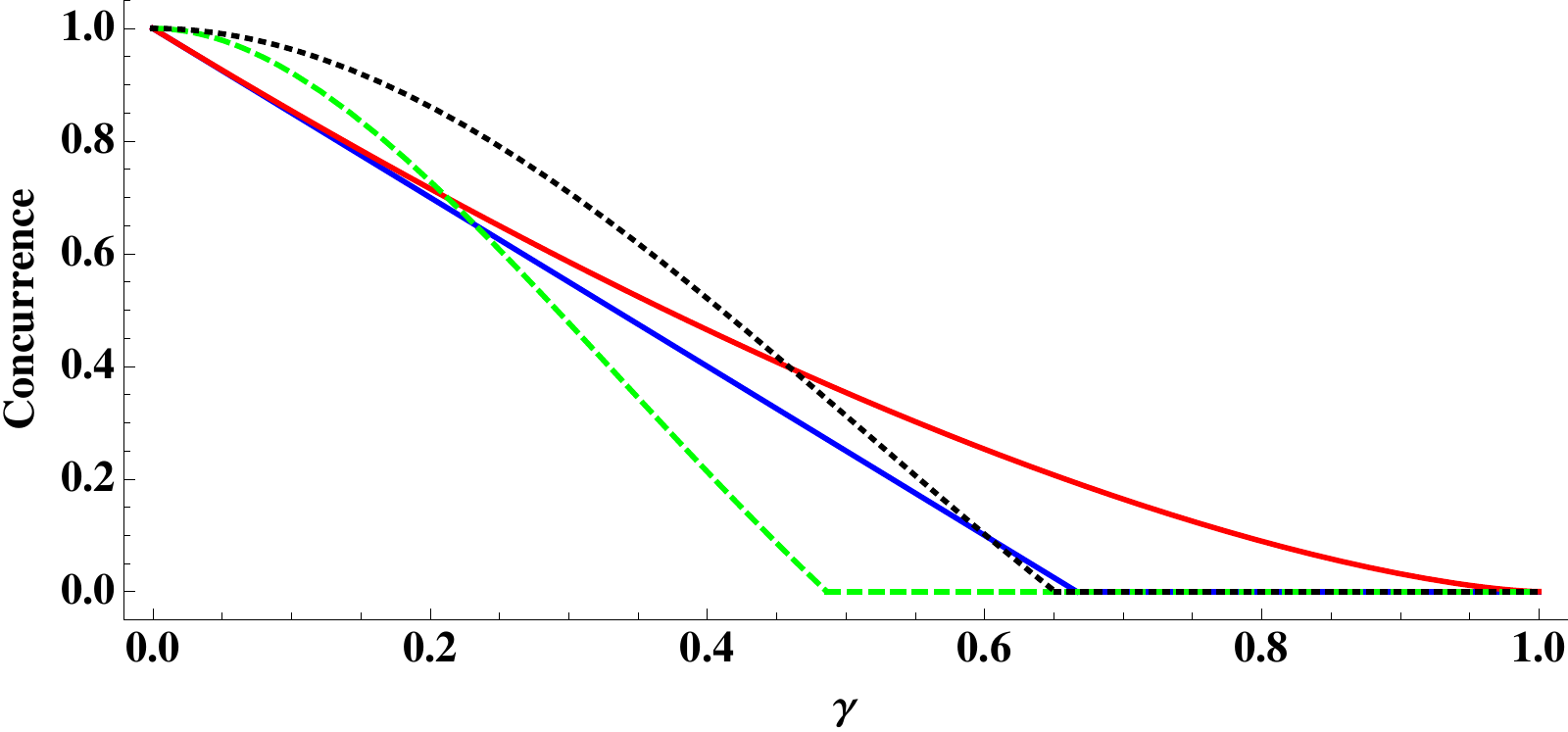} \\
\caption{(Color online) Codeword overlap (top) and concurrence (bottom), as a function of the damping parameter $\gamma$, for different codes: dual-rail, Eq.~(5) (blue), three-qubit repetition, Eq.~(6) (red), bosonic, Eq.~(7) (green, dashed), and four-qubit, Eq.~(8) (black, dotted) \cite{OurFuturePaper}.}
\label{fig:worstcase}
\end{figure}

The results are presented in Figs. 1 and 2. In the former, we observe that the overlaps are zero when the channel approaches perfect transmission: as expected, the orthogonal input codewords remain orthogonal; equally, the entanglement remains maximum. As the losses increase, the overlaps grow; at the same time, the entanglement diminishes. Alone, this fact is highly unsurprising; however, we remark that also the \textit{ordering} established through one measure is reflected on the other. That is to say, the code which offers the best performance, in terms of codeword fidelities, in a given regime, also returns the highest safeguarded entanglement; as the parameters vary, the relative ordering changes as well, and this is observed in both figures of merit. However, the exact point in which a change of ordering occurs is not strictly always the same: between certain encodings, the channel parameters in which a crossing occurs may differ by small amounts. This is particularly noticeable when examining the three-qubit repetition codes, which is highly deformable and suggests that variations may be related to the codeword-dependent skewness.

All of the above suggests a strong link between the overlapping properties of the output alphabet, and the amount of entanglement capable of being safeguarded by means of a given scheme. Nevertheless, we note that no a priori reason exists to suggest that the crossing points should be precisely the same: after all, we are dealing with measures of apparently different character. 

A further observation is in order: while certain encodings lead to entanglement sudden death \cite{eberly,EberlyNew,bjork}, the equivalent catastrophic breakdown in terms of overlaps (average codeword fidelity equalling unity) does not happen in any encoding until the channel becomes fully lossy. This would suggest that the new measure is capable of portraying certain characteristics which would be otherwise lost in an analysis solely based on the entanglement.
\begin{figure}[t!]
\includegraphics[scale=0.5]{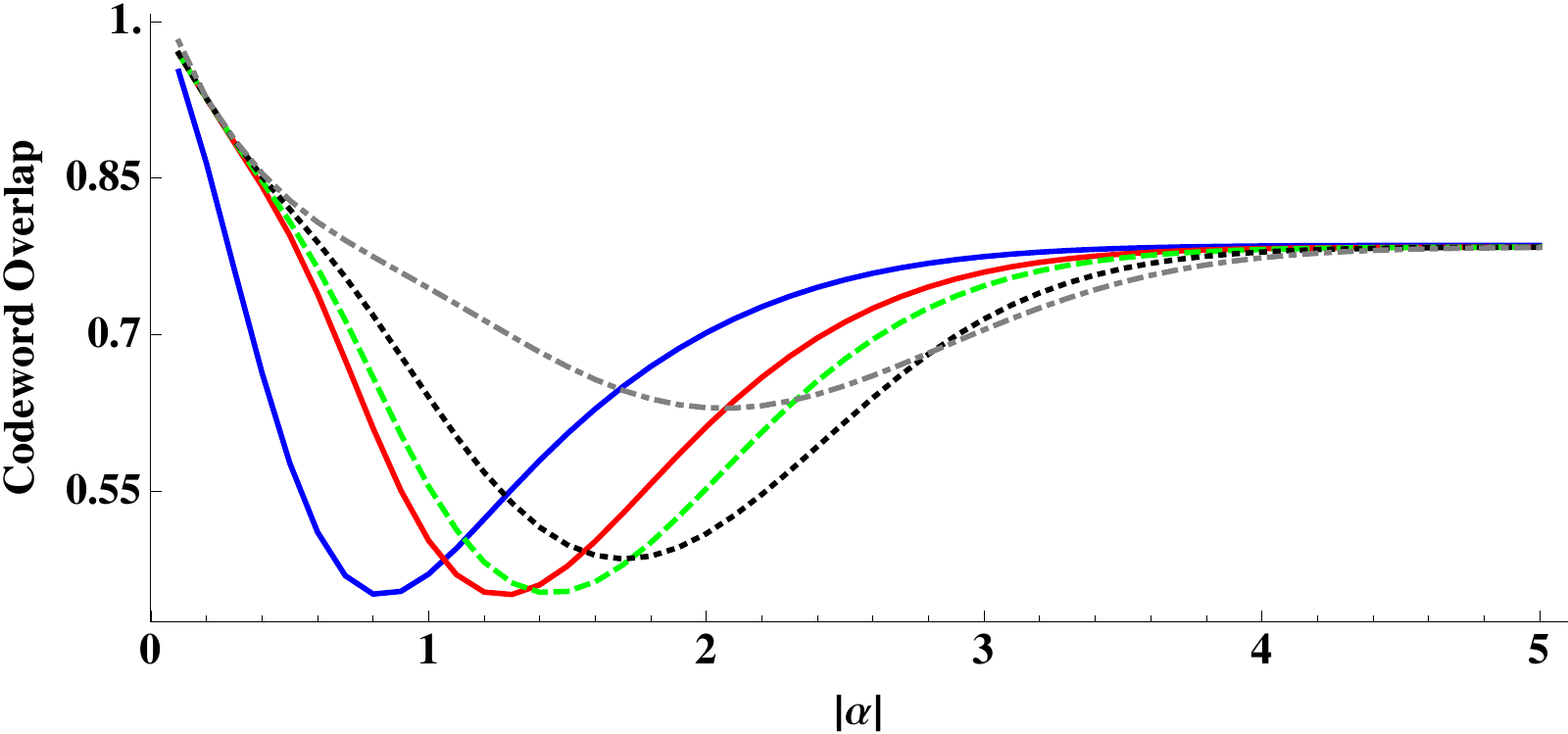} \\
\includegraphics[scale=0.5]{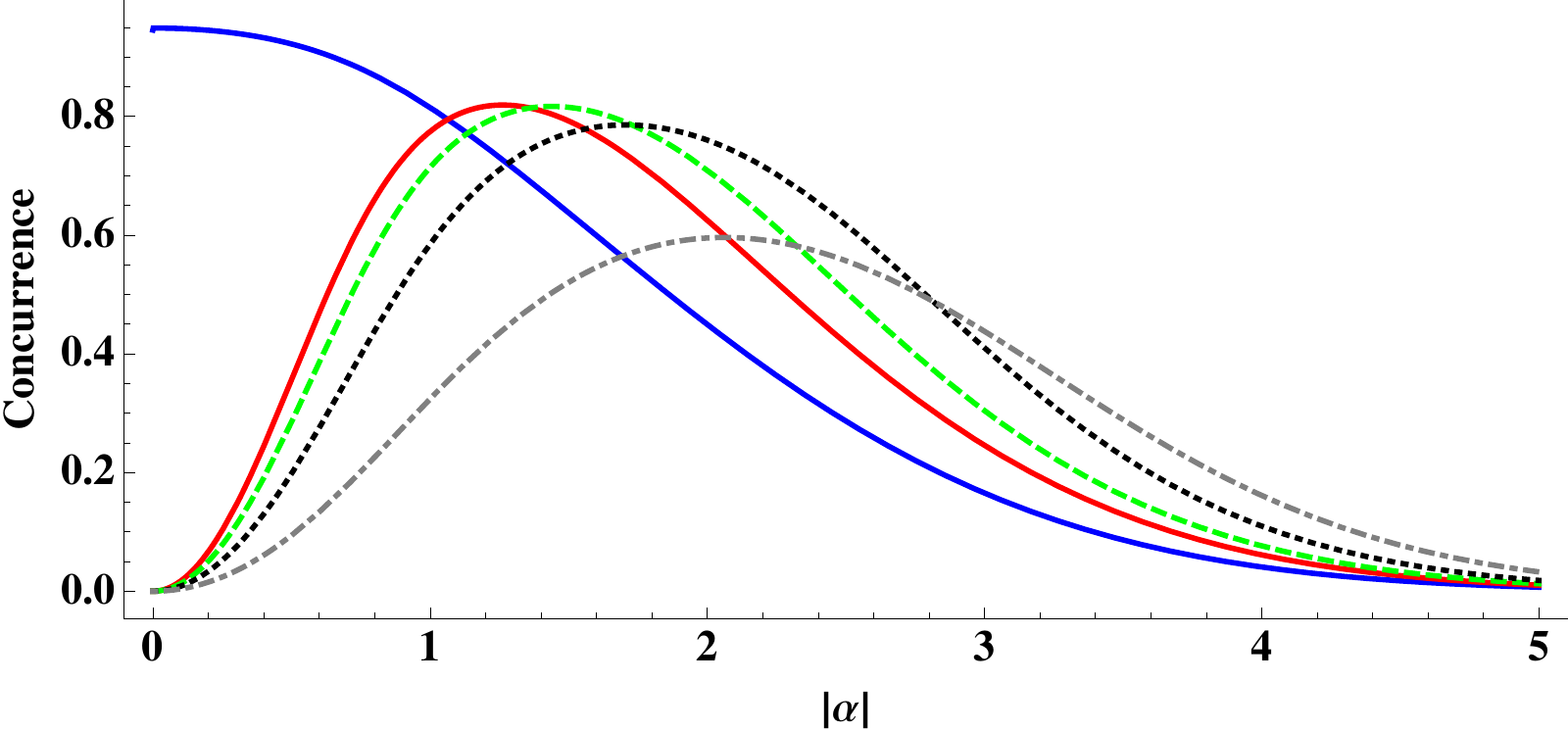}
\figcaption{(Color online) Codeword overlap (top) and concurrence (bottom) for direct transmission Eq.~(9) and the coherent-state code Eq.~(10), after transmission through a $\gamma=0.32$ channel, as a function of the coherent-state superposition size $|\alpha|$. Encodings with 3 (red), 5 (green, dashed), 11 (black, dotted), and 51 (grey, dashed) qubits are depicted in addition to the direct transmission case (blue) \cite{OurFuturePaper} .}
\label{fig:full066}
\end{figure}

In Fig. 2, we observe the behavior of different coherent-state encodings as a function of the superposition size $|\alpha|$. Again, a good agreement is found between the ordering obtained through concurrence and codeword overlaps. Notable is the difference in the case of direct transmission: while it is possible to obtain a maximally entangled state in the limit of $\alpha \rightarrow 0$, the alphabet as a whole becomes indistinguishable and thus impractical for encoding purposes - a feature reflected by the codeword overlap measure.

\section{Gaussian Error Correction No-Go}

Optical modes of the electromagnetic field, whose quadratures satisfy the canonical commutation relations, provide a natural testbed for a wide range of quantum information concepts \cite{PeterReview}. Here, logical information is encoded by means of a truly \textit{continuous} logical alphabet; for instance, the codewords defined by the position eigenstates $\{ |x\rangle_L \}$. In this scenario, the so-called Gaussian operations - defined as those that map Gaussian states into Gaussian states - are of great relevance due to the ease in which they can be implemented with current experimental resources.

Nevertheless, the capabilities of the Gaussian set are not without restrictions. Given the fragility of the quantum resources in face of ubiquitous decoherence mechanisms, a significant limitation is that such operations are incapable of distilling higher entanglement from less entangled Gaussian states \cite{Giedke,Eisert,Fiurasek}, or, in close relation, unable to protect Gaussian states from the widespread class of Gaussian errors \cite{CerfNoGo}.

Interestingly, Gaussian transformations alone suffice to suppress and correct non-Gaussian errors acting on arbitrary states, but in particular, the Gaussian transformations also suffice when the non-Gaussian errors act upon Gaussian input states \cite{BraunsteinQEC, Aoki, niset2, PeterNote}; equally worth noting is the fact that a Gaussian error channel, when acting upon specific non-Gaussian input states, can exhibit stochastic, non-Gaussian behaviour \cite{CMM}, however, in this case Gaussian encoding and decoding procedures alone appear to be incapable of correcting such errors, and non-Gaussian operations should be accounted for \cite{GVR}.

In this section, we'll employ the codeword overlap measure in order to explore what kind of statements it allows us to make in the context of continuous-variable QEC, in particular, in relation to the aforementioned No-Go theorem for all-Gaussian QEC \cite{CerfNoGo}. Most importantly, such an approach enables us to obtain fundamental insights in a rather distinct way, without the need for calculating an entanglement measure (such as the logarithmic negativity employed in Ref.~\cite{CerfNoGo}). We note that the quantity as developed in Eq.~(\ref{fidoverlap}) is based on averaging the fidelity between the channel-output states which originate from two orthogonal input states. However, here, in order to include the possibility of non-orthogonal input states as it is typically the case for a Gaussian alphabet, we'll consider two general states: if the output overlap between two arbitrary states cannot be reduced through QEC, then, in particular, it also holds that the overlap between two orthogonal states (after their channel transmission) will not decrease, and integrating over all possible pairs of states will furthermore not result in a lower quantity, thus establishing the desired result.

The fidelity between two states $\rho_1$ and $\rho_2$ is given, in the characteristic function formalism, by
\begin{align}
F = \frac{1}{\pi^N} \int d^Nx \; \chi_{\rho_1}(x) \; \chi_{\rho_2}(-x) \quad .
\end{align}
For Gaussian states with zero mean, the characteristic function is given by $\chi_{\rho_{i}}=e^{-\frac{1}{2}x^T\sigma_{i}x}$, and we can re-write the above expression as
\begin{align}
F = \frac{1}{\pi^N} \int d^Nx \; e^{-\frac{1}{2} x^T \sigma_1 x} \; e^{-\frac{1}{2} (-x)^T \sigma_2 (-x)} \quad ,
\end{align}
which evaluates to \cite{Scutaru}	
\begin{align}
F = \frac{2}{\sqrt{\Delta + \delta} - \sqrt{\delta}} \quad ,
\label{Forig}
\end{align}
with $\Delta = \det\left(\sigma_1 + \sigma_2\right)$ and $\delta = \left(\det \sigma_1 - 1\right) \left(\det \sigma_2 - 1\right)$.
Now, the action of a Gaussian channel on the level of the covariance matrices is given by $\gamma \rightarrow M \gamma M^{T} + N$ (see Appendix B for details). Taking this transformation in acount, the resulting fidelity is
\begin{align}
F^{\prime} = \frac{2}{\sqrt{\Delta^{\prime} + \delta^{\prime}} - \sqrt{\delta^{\prime}}} \quad ,
\label{Fprime}
\end{align}
where $\Delta^{\prime} = \det\left(M\sigma_1M^{T} + M\sigma_2M^{T} + 2N\right)$, and \quad \quad $\delta^{\prime} = \left(\det (M \sigma_1 M^{T} + N) - 1\right) \left(\det (M \sigma_2 M^{T} + N)- 1\right)$.

Now, to establish a relationship between $F$ and $F^{\prime}$, we must evaluate the resulting effect from $M$ and $N$ in the above (for detailed arguments see Appendix C). We are faced with three relevant cases: (i) $|\det M| = 1$, (ii) $|\det M| > 1$ and (iii) $|\det M| > 1$. The first case is trivially evaluated: when $\det N=0$, this corresponds to a Gaussian unitary, which, as expected, yields exactly the original fidelity, \textit{i.e.}, $F^{\prime}=F$. For $\det N > 0$, representing the addition of classical (thermal) noise, one observes the difference between the square roots in Eq.~(\ref{Fprime}) to diminish, and thus the fidelity to increase. The second case also finds a straightforward solution based on the same argument. The third case, however, contemplates channels which do not necessarily induce a spreading of the Gaussian state, and requires a more careful analysis. In this case, the action of the map results in a ``contraction" towards a common state. This is clearly exemplified by the prototypical amplitude damping channel, using two displaced thermal states as inputs: as the loss parameter $\gamma$ increases, the states are gradually attracted towards the vacuum state; however, the increased purity (and correspondingly reduced values of $\det \sigma_i$) plays no role in diminishing the overlaps.

We note furthermore that displacements (\textit{i.e.}, shifts in the first-order moments) bear no effect in the above, simply rescaling $F$ and $F^{\prime}$ by a fixed amount.

One can then conclude that, except when $\mathcal{E}$ corresponds to a symplectic operation, either the additional noise will cause a spreading of the Gaussian state, or the states will be contracted towards a common state. In any case, the fidelity to any other state subject to the same action is bound to increase. When the decoding operation is equally symplectic, one can at best re-obtain the original state (through $\mathcal{R} = \mathcal{E}^{-1}$), assuming, of course, that the encoding operation already had the best choice of unitary operations.

In the above discussion, eventually we considered input states of a Gaussian nature, thus reproducing the known No-Go result for all-Gaussian QEC \cite{CerfNoGo} in terms of our fidelity-based codeword criterion and independently of entanglement measures.  The treatment of Gaussian codeword pairs above is analogous to our earlier treatment for the qubit codes, and the argumentation follows through depending on the assumption of a Gaussian input alphabet (see Appendix C).

\section{Conclusions and Outlook}

We have defined a quantitative measure based indirectly on the amount by which the well-known quantum error correction criteria by Knill and Laflamme are violated, proposing the use of averaged codeword fidelities as a figure of merit to evaluate different quantum protocols. By quantifying how distinguishable originally orthogonal inputs emerge from an error channel, the measure finds a natural interpretation as a translation, into the quantum regime, of the classical coding theory notion of \textit{confusability} of an alphabet \cite{Shannon,*Lovasz}.


While initially harder to compute, by requiring the evaluation of the fidelity between two generally mixed states, the codeword overlap measure does not require the optimization (minimization) necessary to obtain the worst-case fidelity. At the same time, it accurately depicts the alphabet's distinguishability behaviour which fails to be portrayed by means of the entanglement fidelity or conventional entanglement measures. We emphasize that properly describing this distinguishability, both in an ideal and in a noisy or lossy quantum channel, becomes particularly important when the ideal (quantum) codewords are already non-orthogonal, as it is often the case when the quantum information carriers are continuous-variable oscillator states.

The employment of the codeword overlap reflects \textit{qualitatively} the behaviour found when quantifying the performance by means of other figures of merit, in particular here the concurrence. In other words, whenever the variation of a certain channel parameter causes one scheme to improve (reduce) the average codeword fidelity in comparison to another scheme, one also observes that the first scheme will result in an improved figure in the safeguarded entanglement. This matching was also found to be \textit{quantitatively} exact for certain choices of codes, however, the precise reasons for such behaviour are hitherto unknown.




We have also employed the new measure to reaffirm the impossibility of improving the transmission of Gaussian states, subject to Gaussian noisy channels, through Gaussian operations alone. In this case, the No-Go result was found independently of an entanglement measure, solely based upon the overlap of Gaussian codeword pairs.


Finally, since the measure is, in principle, accessible and computable in those instances where entanglement is found hard to quantify (\textit{i.e.}, infinite-dimensional non-Gaussian states), we expect these results to shed light in other aspects of (photonic) error correction in particular and quantum information in general.

$\quad$ \\

\section*{Acknowledgements}
The authors thank Carlo Cafaro for comments to the manuscript and acknowledge discussions with Dominik H\"orndlein in the early stages of this project. Portions of this work were carried out while R.W. was visiting the Institute for Quantum Computing in Waterloo, Canada. He is grateful for the hospitality and inspiring discussions with the Optical Quantum Communication Theory group. Financial assistance from the ``Collaborative Training in Quantum Information Processing" program is gratefully acknowledged. This work was supported by the German Research Foundation (Deutsche Forschungsgemeinschaft - DFG) via its Emmy Noether Program, and by the Federal Ministry of Education and Research (Bundesministerium f\"ur Bildung und Forschung - BMBF) by means of the HIPERCOM project.

\bibliography{losbenches}

\appendix

\section{Calculating the concurrence and average codeword fidelity}

In obtaining the codeword overlap curves depicted in Figs. (1) and (2), the following procedure is used:

For each encoding, an input qubit $|Q\rangle$ is prepared in the state $\cos \frac{w}{2}|{0}\rangle_L + e^{i \theta}\sin \frac{w}{2}|{1}\rangle_L$; concomitantly, a second state is prepared, $|\tilde{Q}\rangle = \sin \frac{w}{2}|{0}\rangle_L - e^{i \theta} \cos \frac{w}{2}|{1}\rangle_L$, ensuring that $\langle \tilde{Q} | Q \rangle = 0$. Each of those states is then subject to the transmission channel characterized by the operators in Eq.~(\ref{amplitudedampingkrausoperators}). Followed by each code's respective decoding procedure, the following outputs are found:

Direct transmission results in
\begin{align}
\label{amplitudedampingrho1}
&\rho_{Q,direct} = \nonumber \\
&\frac{1}{2}
\left(
\begin{array}{cc}
1+ \gamma + \cos w - \gamma \cos w & e^{-i \theta} \sqrt{1-\gamma} \sin w \\
e^{i \theta} \sqrt{1-\gamma} \sin w & (\gamma - 1) (\cos w -1)
\end{array}
\right)
\end{align}
and
\begin{align}
\label{amplitudedampingrho2}
&\rho_{\tilde{Q},direct} = \nonumber \\
&\frac{1}{2}
\left(
\begin{array}{cc}
1+ \gamma + \cos w + (1- \gamma) \cos w & -e^{-i \theta} \sqrt{1-\gamma} \sin w \\
-e^{i \theta} \sqrt{1-\gamma} \sin w & -(\gamma-1)(\cos w + 1)
\end{array}
\right) \; .
\end{align}
\begin{align}
&\mbox{To the dual-rail encoding corresponds} \nonumber \\
\label{dr-output1}
& \rho_{Q,dual-rail} = \nonumber \\
& \frac{1}{2}
\left(
\begin{array}{cc}
   (1+(-1+\gamma ) \cos w) &   e^{-i \theta } (-1+\gamma ) \sin w \\
   e^{i \theta } (-1+\gamma ) \sin w &   (1+\cos w-\gamma  \cos w) \\
\end{array}
\right)
\end{align}
and
\begin{align}
\label{dr-output2}
& \rho_{\tilde{Q},dual-rail} = \nonumber \\
&\frac{1}{2}
\left(
\begin{array}{cc}
   (1+\cos w-\gamma  \cos w) & -  e^{-i \theta } (-1+\gamma ) \sin w \\
 -  e^{i \theta } (-1+\gamma ) \sin w &   (1+(-1+\gamma ) \cos w) \\
\end{array}
\right) \; .
\end{align}

\begin{widetext}
The 3-qubit code produces
\begin{align}
\label{amplitudedampingrho3}
\rho_{Q,3-qubit} &=
\frac{1}{2}
\left(
\begin{array}{cc}
1+(3-2 p) p^2+(p-1)^2 (1+2 p) \cos w &  e^{-i \theta } (1-p)^{3/2} \sin w \\
e^{i \theta } (1-p)^{3/2} \sin w  & (p-1)^2 (2+4 p) \sin^2 \frac{w}{2}
\end{array}
\right)
\end{align}
and
\begin{align}
\label{amplitudedampingrho4}
\rho_{\tilde{Q},3-qubit} &=
\frac{1}{2}
\left(
\begin{array}{cc}
1+(3-2 p) p^2-(p-1)^2 (1+2 p) \cos w & -e^{-i \theta } (1-p)^{3/2} \sin w \\
 -e^{i \theta } (1-p)^{3/2} \sin w & (p-1)^2 (1+2 p) (1+\cos w)
\end{array}
\right) \; .
\end{align}

For the bosonic encoding, one finds
\begin{align}
\label{bosonicrho1}
\rho_{Q,bosonic} &=
\frac{1}{2}
\left(
\begin{array}{cc}
1-(\gamma -1)^3 (1+3 \gamma ) \cos w & - e^{-i \theta } (\gamma -1)^3 (1+3 \gamma ) \sin w \\
 - e^{i \theta } (\gamma -1)^3 (1+3 \gamma ) \sin w &  1+(\gamma -1)^3 (1+3 \gamma ) \cos w \\
\end{array}
\right)
\end{align}
and
\begin{align}
\label{bosonicrho2}
\rho_{\tilde{Q},bosonic} &=
\frac{1}{2}
\left(
\begin{array}{cc}
1+(\gamma -1)^3 (1+3 \gamma ) \cos w &  e^{-i \theta } (\gamma -1)^3 (1+3 \gamma ) \sin w \\
  e^{i \theta } (\gamma -1)^3 (1+3 \gamma ) \sin w & 1-(\gamma -1)^3 (1+3 \gamma ) \cos w \\
\end{array}
\right) \; .
\end{align}

Finally, for the ``approximate" encoding, the outputs are
\begin{align}
\label{bjorkrho1}
&\rho_{Q,approximate} = \nonumber \\
&\left(s
\begin{array}{cc}
 \frac{1}{2} \left( 1+\gamma ^2 (2 \gamma -1)+(\gamma-1 )^2 (1+2 \gamma ) \cos w \right) & \frac{1}{4} e^{i \theta } \left(\gamma ^2-\gamma ^3+e^{-2 i \theta } \left(2+\gamma ^2 (3 \gamma -5)\right)\right) \sin w \\
  \frac{1}{4} e^{-i \theta } \left(\gamma ^2-\gamma ^3+e^{2 i \theta } \left(2+\gamma ^2 (3 \gamma -5)\right)\right) \sin w &
\frac{1}{2} \left( 1+\gamma ^2-2 \gamma ^3-(\gamma -1 )^2 (1+2 \gamma ) \cos w \right) \\
\end{array}
\right)
\end{align}
and
\begin{align}
\label{bjorkrho2}
&\rho_{\tilde{Q},approximate} = \nonumber \\
&\left(
\begin{array}{cc}
 \frac{1}{2} \left(1+\gamma ^2 (-1+2 \gamma )-(-1+\gamma )^2 (1+2 \gamma ) \cos w\right) & \frac{1}{4} e^{-i \theta } (-1+\gamma ) \left(2+\gamma
 \left(2+\left(-3+e^{2 i \theta }\right) \gamma \right)\right) \sin w \\
 \frac{1}{4} e^{-i \theta } \left((-1+\gamma ) \gamma ^2+e^{2 i \theta } \left(-2+(5-3 \gamma ) \gamma ^2\right)\right) \sin w & \frac{1}{2}
\left(1+\gamma ^2-2 \gamma ^3+(-1+\gamma )^2 (1+2 \gamma ) \cos w\right) \\
\end{array}
\right) .
\end{align}
\end{widetext}

With the above, Eq.~(\ref{fidoverlap}) is then computed by means of a numeric integration procedure over $w$ and $\theta$. At each value of $\gamma$, approximately 1000 states are employed to produce the average, although for less deformable encodings a smaller sample already resulted in an adequate agreement with the asymptotic behaviour.

For evaluating the entanglement-safeguarding capabilities of a scheme, the maximally entangled state $\frac{1}{\sqrt{2}}\left( |0\rangle |0\rangle_L + |1\rangle |1\rangle_L \right)$ is employed, transmitting the encoded mode through the lossy channel, afterwards followed by the decoding procedures. We note that the decoding operations reduce the encoded state back to a qubit subspace, therefore allowing the entanglement to be computed on a $2\times2$ Hilbert space.

For the entanglement analysis developed in the text, we consider Wooters' concurrence, which is obtained through
\begin{equation}
\label{conceq}
C = \max\{0,\sqrt{\lambda_1}-\sqrt{\lambda_2}-\sqrt{\lambda_3}-\sqrt{\lambda_4} \} \; .
\end{equation}
Here, $\lambda_i$ are the eigenvalues, in decreasing order, of $\rho \tilde{\rho}$, where $\tilde{\rho} = ( \sigma_{y,1} \otimes \sigma_{y,2} ) \rho^{*} ( \sigma_{y,1} \otimes \sigma_{y,2} )$, and $\sigma_{y,i}$ is the Pauli $Y$ operator in the $i$-th mode.

In the case of coherent-state encodings evaluated in Fig. (2), the input state for both direct transmission and the different encodings is
\begin{equation}
|Q\rangle = \frac{1}{\sqrt{N(\alpha)}} ( \sqrt{w} |-\alpha\rangle + e^{i \theta} \sqrt{1-w} |\alpha\rangle ) \quad ,
\end{equation}
with $0 \geq w \geq 1$, $0 \geq \theta \geq \pi$, and the normalization constant $N(\alpha) = 1 + 2 \cos \theta \sqrt{w(1-w)} e^{-2|\alpha|^2}(ab^*+a^*b)$, required due to the non-orthogonal nature of the coherent-state alphabet. When considering the error correction protocol \cite{GVR}, the deterministic operation of the scheme is considered; in other words, the transformation in Eq.~(\ref{coherentencoding}) is only obtained for large superposition sizes, with smaller $|\alpha|$ incurring an erroneous component (see ref. \cite{OurFuturePaper} for details). This induces in a double-tradeoff between the superposition size and the figure of merit being considered: while on the one hand, larger ``cat states" are desired - not only due to their inherenty larger distinguishability, but also for the operation of the non-Gaussian Hadamard gates found in the protocol - on the other hand, this effects an increased channel-induced phase-flip probability.

For computing the concurrence in the later case, two-mode maximally entangled states are employed,
\begin{equation}
\label{maxentstate}
|\Phi^-\rangle = \frac{1}{\sqrt{2-2e^{-4|\alpha|^2}}} (|\alpha,\alpha\rangle - |-\alpha,-\alpha\rangle) \quad .
\end{equation}
Again, the first mode is kept while the second is sent through the error-correcting scheme.  Finally, for the above calculations, it helps to express the density matrices in terms of an orthogonal basis $\{ |u_{\alpha}\rangle,|v_{\alpha}\rangle \}$, such that
\begin{align}
|\alpha\rangle = \mu_\alpha |u_{\alpha}\rangle& + \nu_\alpha |v_{\alpha}\rangle \\
|-\alpha\rangle = \mu_\alpha |u_{\alpha}\rangle& - \nu_\alpha |v_{\alpha}\rangle \nonumber \\
\mbox{with } \mu_\alpha = \left(\frac{1+e^{-2|\alpha|^2}}{2}\right)^{\frac{1}{2}}& \; \textrm{and} \; \nu_\alpha = \left(\frac{1-e^{-2|\alpha|^2}}{2}\right)^{\frac{1}{2}} \; .\nonumber
\end{align}

\section{Gaussian Formalism}

\subsection{Gaussian States}

Here we briefly review the Gaussian formalism, adopting the convention from \cite{Giedke}.

Given the Weyl operators
\begin{align}
W(x) = e^{-ix^TR},
\end{align}
where $x \in \mathbb{R}^{2n}$ and $R=(X_1,P_1,..,X_n,P_n)^T$, with the commutator relations $[X_j,P_k]=i\delta_{jk}$, a Gaussian state $\rho$ is defined as having a Gaussian characteristic function $\chi_{\rho}(x) = \mbox{tr}[{\rho}W(x)]$. An equivalent definition can be given in terms of the state's Wigner function, which can be obtained from the characteristic function by a Fourier transform, and is also Gaussian-shaped. One defines the first and second order moments $\textbf{d}$ and $\gamma$, respectively the displacement vector and the covariance matrix (CM), by $d_i = \langle x_i \rangle$ and $\gamma_{ij} = \langle x_i x_j + x_j x_i\rangle - 2 d_i d_j$. In most cases, $\textbf{d}$ can be set to zero without loss of generality. The CM satisfies $\gamma = \gamma^T \geq i J_n$, where we define the symplectic matrix 
\begin{align}
J_n = \bigoplus_{k=1}^{n} J_1 \, , \quad \quad J_1 = \left(
\begin{array}{cr}
0 & -1 \\
1 & 0 \end{array} \right) .
\end{align}
The displacement vector and covariance matrix completely determine the state $\rho$, whose density operator can be written as
\begin{align}
\rho = \pi^{-n} \int_{\mathbb{R}^{2n}} dx \, e^{ - \frac{1}{4}x^T \gamma x + i d^{T}x} W(x) \; .
\label{opform}
\end{align}

Of special relevance is the maximally entangled state (MES), corresponding to an infinitely squeezed two-mode squeezed state (TMSS), with CM
\begin{align}
\lim_{r \rightarrow \infty}\gamma(r) = \left(
\begin{array}{cc}
A_r & C_r \\
C_r & A_r \end{array} \right) ,
\label{TMSS}
\end{align}
where $A_r = \cosh r \mathbb{1}$ and $C_r = \sinh r \Lambda$ are both $2n \times 2n$ matrices, and
\begin{align}
\Lambda = \mbox{diag} \left(1, -1, 1, -1, ... , 1, -1\right) .
\end{align}

\subsection{Gaussian Operations}

A Gaussian channel is defined as a map $ \mathcal{E} $ taking Gaussian states into Gaussian states, cf. $\rho^{\prime}=\mathcal{E}(\rho)$. Following the Choi-Jamiolkowski isomorphism between completely positive maps and positive operators \cite{choi}, to every Gaussian map $\mathcal{E}$ there corresponds an operator $\hat{E}$,
\begin{align}
\hat{E}_{12} = \lim_{r\rightarrow\infty} \left(\mathcal{E} \otimes \mathbb{1} \right) \left( |\phi \rangle_{12} \langle \phi | \right) \: ,
\label{choieq}
\end{align}
this equation allowing us to re-interpret the transmission of an arbitrary state $\rho$ through $\mathcal{E}$ as a teleportation using $\hat{E}_{12}$ as the entangled resource state, \textit{i.e.},
\begin{align}
\mathcal{E}(\rho) \propto \mbox{tr}_2[\hat{E}_{12}^{T_2}\rho_2] = \mbox{tr}_{23} (\hat{E}_{12}\rho_{3} |\phi\rangle_{23} \langle \phi | ) .
\label{eqdotraco}
\end{align}
One should note that, since $\mathcal{E}$ maps Gaussian states into Gaussian states, and $|\phi\rangle$ in Eq.~(\ref{choieq}) can be taken as the limit of a Gaussian state, $\hat{E}$ must itself correspond to a Gaussian operator which, similarly to (\ref{opform}), can also be written as
\begin{align}
\hat{E} = \int_{\mathbb{R}^{2n}} dx \, e^{ - \frac{1}{4}x^T \Gamma x + i D^{T}x - C} W(x) ,
\label{opform2}
\end{align}
with appropriately-defined CM $\Gamma$, displacement vector $D$ and a normalization constant $C$.  Now, by employing Eq.~(\ref{eqdotraco}) and replacing $\hat{E}_{12}$ with the operator in Eq.~(\ref{opform2}), one can obtain the action of $\mathcal{E}$ on a general state, ie, obtain $\gamma^{\prime}$ and $d^{\prime}$ in $\mathcal{E}: \rho_{\gamma,d} \rightarrow \rho_{\gamma^{\prime},d^{\prime}}$ from $\Gamma$ and $D$ \footnote{The interested reader should follow sections II and III in ref. \cite{Giedke} for a complete discussion of this procedure.}.

We now consider the set of operations which can be implemented by augmenting our system with additional (Gaussian) ancillary states, performing Gaussian unitary operations over the whole combined system and discarding (tracing over) the ancillas, thus obtaining the class of Gaussian completely positive trace-preserving (CPTP) maps. The action of such a map on a state with CM $\gamma$ is given by
\begin{align}
\gamma \rightarrow M \gamma M^{T} + N \: ,
\label{MgammaM}
\end{align}
with $M$ real and $N \geq 0$ real and symmetric. We must also have 
\begin{align}
\det N \geq (\det M -1 )^2 ,
\label{relNM}
\end{align}
lest the complete-positivity requirement be violated. The Gaussian operator corresponding to this operation has the CM
\begin{align}
\Gamma = \lim_{r \rightarrow \infty} \left(
\begin{array}{cc}
M^{T} A_r M + N & M^{T} C_r \\
C_r M & A_r \end{array} \right) .
\label{MAM}
\end{align}

The solutions to Eq.~(\ref{relNM}) with $N=0$ (adding no extra noise) and $\det M = 1$ (preserving the sum of areas) are dubbed symplectic transformations. In the quantum-optical context, these are exemplified by unitaries such as squeezers, phase shifters, or lossless beam-splitters. One can easily verify those to be the channels with minimal entanglement degradation \cite{CerfNoGo}, or equally, those preserving the newly-developed codeword overlap measure.

Finally, we note that the identity map, $\mathcal{I}$, is obtained in Eq.~(\ref{MgammaM}) by taking $M = \mathbb{1}$ and $N=0$ (being thus a symplectic operation), and reduces the expression (\ref{MAM}) to the CM of the maximally entangled ($r \rightarrow \infty$) TMSS, Eq.~(\ref{TMSS}). 

\section{Fidelity between Gaussian states}

The effect of a Gaussian channel on the level of the covariance matrices is given by $\gamma \rightarrow M \gamma M^{T} + N$. In order to facilitate the treatment of such expression when evaluating fidelities, we note that the above Gaussian channel can be parametrized by means of a singular value decomposition. This gives rise to an equivalent channel characterised by the matrices
\begin{align}
\label{symplecticdecomp}
M' &= S V M U \; \quad \; \mbox{ and} \\
N' &= S V N V^T S \nonumber
\end{align}
Since $U$, $V$ and $S$ are all symplectic, the overlaps between the original and transformed channel remain unaltered. Now, $\det M' = \det M$ (and equally for $N'$ and $N$); furthermore, without loss of generality we can choose $M$ as proportional to the identity, i.e.,  $M \propto \eta \mathbb{1}$. Doing so greatly simplifies the expression for the fidelity after the Gaussian operations, since one can then basically employ the original determinants, up to a scaling factor.

Now, to establish a relationship between $F$ and $F^{\prime}$, we must evaluate the resulting effect from $M$ and $N$ in the above. We are faced with three relevant cases: (i) $|\det M| = 1$, (ii) $|\det M| > 1$ and (iii) $|\det M| > 1$.

The first case, $|\det M| = 1$, with $\det N=0$ corresponds to a Gaussian unitary and is trivially evaluated, yielding, as expected, exactly the original fidelity

For $\det N > 0$, representing the addition of classical (thermal) noise, and equivalently for the second case, $|\det M| > 1$, with help of the parametrization in Eq.~(\ref{symplecticdecomp}), one trivially observes the difference between the square roots in Eq.~(\ref{Fprime}) to diminish, and thus the fidelity to increase. The later holds for quantum-limited maps \cite{Solomon} where $N$ bounds Eq.~(\ref{relNM}); and evidently for those cases where a second, (classical) noise channel follows.

The third and final case, however, involves a more elaborate analysis. Using the expression for the symplectic transformation Eq.~(\ref{symplecticdecomp}) in Eq.~(\ref{Fprime}), expressing the relevant quantities in terms of the parameter $\eta$ governing $M$ and $N$, and employing the physicality constraints for Gaussian channels (see Eq.~(\ref{relNM}) in Appendix B), one finds, after long but otherwise straightforward calculations, that
\begin{align}
\label{thebigresult}
F - F^{\prime} > 0 \;
\end{align}
has an empty solution set; and thus, the fidelity is indeed bound to increase.

Finally, in order to fully appreciate that the treatment with codeword pairs of Gaussian states developed in section IV is analogous to our earlier treatment for the qubit codes in section III, and also that the argument follows through depending on the assumption of a Gaussian input alphabet, notice the following. A pair of continuous-variable codewords may be written as
$|\psi_1\rangle = \int \, dx \, \psi_1(x)\,|x\rangle_L$ and
$|\psi_2\rangle = \int \, dx \, \psi_2(x)\,|x\rangle_L$, similar to our $|Q\rangle$ and
$|\tilde Q\rangle$ states for qubits, but with the general Bloch-sphere parameters (App. A) this time
replaced by two general wave functions $\psi_1(x)$ and $\psi_2(x)$; the states $|\psi_1\rangle$ and
$|\psi_2\rangle$ are also no longer restricted to be orthogonal. Now assuming that the logical states
$|x\rangle_L$ that span the continuous codespace are the result of an arbitrary, unitary Gaussian
encoding operation acting upon a one-mode basis state $|x\rangle$ together with an arbitrary Gaussian multi-mode
ancilla state $|{\rm Gaussian}\rangle$,
$|x\rangle_L = \hat U_{\rm Gaussian} \left( |x\rangle\otimes |{\rm Gaussian}\rangle \right)$, we obtain for the
two input states,
\begin{eqnarray}
|\psi_i\rangle &=& \int \, dx \, \psi_i(x)\,|x\rangle_L \\
&=& \int \, dx \, \psi_i(x)\,\hat U_{\rm Gaussian} \left( |x\rangle\otimes |{\rm Gaussian}\rangle \right)
\nonumber\\
&=& \hat U_{\rm Gaussian} \,\left( \int \, dx \, \psi_i(x) |x\rangle\otimes |{\rm Gaussian}\rangle \right)\,
\nonumber
\end{eqnarray}
with $i=1,2$.
Under the assumption of two Gaussian wave functions $\psi_1(x)$ and $\psi_2(x)$,
it is guaranteed that the resulting state above is again a Gaussian state.
However, for two arbitrary wave functions $\psi_1(x)$ and $\psi_2(x)$, the resulting states
can be non-Gaussian, and the averaging for obtaining our overlap measure would have to
include non-Gaussian codewords as well (corresponding to the more general scenario
of arbitrary signal states subject to Gaussian encoding/decoding operations and
Gaussian error channels). In this latter case, the above analysis for
the Gaussian overlaps would no longer suffice. Therefore, we must assume that
$\psi_1(x)$ and $\psi_2(x)$ represent {\it any} pair of {\it Gaussian} wave functions,
and the above arguments follow through. Note that this conclusion is robust against
a change of the basis of the original signal state, $|x\rangle\rightarrow |n\rangle$
(with $|n\rangle$ being the photon number basis),
in which case
\begin{eqnarray}
|\psi_i\rangle &=& \sum_n c_{i,n}\,|n\rangle_L \\
&=&
\sum_n c_{i,n}\,\hat U_{\rm Gaussian}\left( |n\rangle\otimes |{\rm Gaussian}\rangle \right)\nonumber\\
&=&
\hat U_{\rm Gaussian}\left( \sum_n c_{i,n}\, |n\rangle\otimes |{\rm Gaussian}\rangle \right)\,,
\nonumber
\end{eqnarray}
and where $\hat U_{\rm Gaussian}\left( |n\rangle\otimes |{\rm Gaussian}\rangle \right)$
is generally a non-Gaussian state, but
$\hat U_{\rm Gaussian}\left( \sum_n c_{i,n}\, |n\rangle\otimes |{\rm Gaussian}\rangle \right)$
is a Gaussian state, provided that $|\psi_i\rangle = \sum_n c_{i,n}\,|n\rangle$ is a Gaussian state too.


\end{document}